\documentclass[preprintnumbers,nofootinbib,showpacs,showkeys]{revtex4}
\usepackage{amsmath} \usepackage{graphicx} \usepackage{amsfonts}
\usepackage{array} \usepackage{amsthm} \usepackage{bm}
\usepackage{supertabular}
\usepackage{latexsym}

\usepackage[breaklinks]{hyperref}
\usepackage{color}
\def\n{\noindent}

\begin{document}
\begin{flushleft}
\footnotesize{Published in: Monatshefte f\"{u}r Chemie \textbf{136}, 2017-2027 (2005)\\
DOI: 10.1007/s00706-005-0370-3}
\end{flushleft}

\title{Low-temperature data for carbon dioxide}

\author{Mustapha Azreg-A\"{\i}nou}
\affiliation{Ba\c{s}kent University, Engineering Faculty, Ba\u{g}l\i ca Campus, Ankara, Turkey}

%\date{}

\begin{abstract}
We investigate the empirical data for the
vapor pressure (154$ \leq$$T$$\leq$196 K) and heat capacity
(15.52$ \leq$$T$$\leq$189.78 K) of the solid carbon dioxide. The
approach is both theoretical and numerical, using a computer
algebra system (CAS). From the latter point of view, we have
adopted a cubic piecewise polynomial representation for the heat
capacity and reached an excellent agreement between the available
empirical data and the evaluated one. Furthermore, we have
obtained values for the vapor pressure and heat of sublimation at
temperatures below 195 right down to 0 K. The key prerequisites
are the: 1) Determination of the heat of sublimation of 26250
J$\cdot$mol\textsuperscript{-1} at vanishing temperature and 2)
Elaboration of a `linearized' vapor pressure equation that
includes all the relevant properties of the gaseous and solid
phases. It is shown that: 1) The empirical vapor pressure equation
derived by Giauque \& Egan remains valid below the assumed lower
limit of 154 K (similar argument holds for Antoine's equation), 2)
The heat of sublimation reaches its maximum value of 27211
J$\cdot$mol\textsuperscript{-1} at 58.829 K and 3) The vapor
behaves as a (polyatomic) ideal gas for temperatures below 150 K.
\end{abstract}


\maketitle

\section{Introduction}

Because of the intensive use of
carbon dioxide in industry and research \cite{carbon}, it has
become necessary to determine its thermodynamic, physical and
chemical properties on an extended range of temperatures.
Significant effort has been deployed to build up a database
through observations and theoretical calculations
\cite{Meyers,Giauque,d2,Suzuki,Schnepp,Ron,G,Gray,nist,tool,Jr,IoP}.
From the former point of view, we mention the case of the accurate
measurements due to Giauque \& Egan \cite{Giauque} and from the
latter point of view, the derivation based on the classical
version of the theory of lattice dynamics, which predicts the heat
capacity of carbon dioxide in the range of temperatures
15$<$$T$$<$50 K \cite{Suzuki}, is in a very good agreement with
that obtained through observations \cite{Giauque}.

However, such a good agreement is still out of reach for some
other properties of carbon dioxide due to difficulties from both
experimental and theoretical points of view. For instance, the
empirical determination of the latent heat of sublimation at low
temperatures remains a major obstacle because of the difficulty in
eliminating the superheating of the gas \cite{Giauque}. Similarly,
by way of example, the lagrangian classical treatment of the
two-dimensional rigid rotor is intractable and the theoretical
determination of the heat capacity, mentioned above, had been made
possible at only sufficiently low temperatures ($T$$<$50 K) when
the harmonic approximation is valid \cite{Schnepp}. With that
said, much work has to be done in order to determine further
properties of carbon dioxide particularly at low temperatures,
such properties are still missing in the best compendia.

We will exploit the data available in \cite{Giauque}, which we
refer to as G\&E, and show that it is possible to evaluate the
heat of sublimation $L$ and vapor pressure $p$ at temperatures
5$\leq$$T$$\leq$195 K. A key prerequisite is the determination of
the heat of sublimation at $T$=0 K ($L(0)$=$\epsilon_{0}$). Stull
calculated an average value of $L$ by the method of least squares
using the vapor pressure data measured by different workers
\cite{d2} and obtained a value of 26.3
kJ$\cdot$mol\textsuperscript{-1} (=6286
cal$\cdot$mol\textsuperscript{-1}) for 139$\leq$$T$$\leq$195 K
\cite{nist}. However, the literature citations listed in \cite{d2}
show that Stull did not extract data from G\&E, which is even more
accurate and includes data concerning the heat capacity of the
solid carbon dioxide and other data that could be used to obtain
$L$ at different temperatures. By contrast, G\&E have evaluated
$L$ at 194.67 K using partly their measured data and available
data for $L$ at lower temperatures \cite{Meyers}. They evaluated
the integral of the heat capacity of the solid (change in the
enthalpy) graphically from a smooth curve through their measured
data and obtained a value for $L$ that is merely 10
cal$\cdot$mol\textsuperscript{-1} higher than their measured value
$L_{\rm{meas}}(194.67)$=6030$\pm$5
cal$\cdot$mol\textsuperscript{-1} (25230$\pm$21
J$\cdot$mol\textsuperscript{-1}). They also evaluated the
entropies of the gas and solid at 194.67 K and reached an
\textsl{excellent} agreement between experimental data and
statistics (the experimental \& spectroscopic values of the
entropy of the gas $s_{g}$ they obtained were 47.59 \& 47.55
cal$\cdot$K\textsuperscript{-1}$\cdot$mol\textsuperscript{-1},
respectively, constituting a proof of the third law \cite{d3}).
However, this cumbersome procedure had prevented them from
carrying out a systematic evaluation of the latent heat and
entropy at temperatures covering the range of their measured data.
Furthermore, this procedure (the graphical evaluation) adds a
human error, which is an unknown factor.

In this paper we will carry out a systematic evaluation of the
fore-mentioned physical quantities on a more extended range of
temperatures than that of G\&E using 1) a computer algebra system
(CAS), which eliminates the human error and allows an excellent
adjustment of the parameters in order to achieve a better
accuracy, as well as 2) an established formula for the vapor
pressure. It will be shown below that our reevaluated value of
$L(194.67)$ is 6030.4 cal$\cdot$mol\textsuperscript{-1} (25231
J$\cdot$mol\textsuperscript{-1}). The data for the relevant
quantities will be tabulated at temperatures incremented by 5 K
and plotted. Moreover, the generating codes will be provided,
which allow the evaluation of any quantity at any given
temperature within minutes of time. In this work, we will be
relying on measured data by different workers and on some
empirical formulas derived by graphical interpolation. Since some
of these data are provided without accuracy and some other lack
accuracy due to personal error, it will be difficult to assign
accuracy to our results, as is the case in most compendia. Some
values of $p$ (in Torr) will be given with one significant digit
while other values with 2 or 3 significant digits. The values of
$L$ (of the order of 26000 J$\cdot$mol\textsuperscript{-1}) will
be given with five digits without decimals, assuming an error not
higher than 0.35\%. The accuracy of the results for $p$ and $L$
can be read by comparing with the available measured data.

\begin{figure}[h]
\centering
  \includegraphics[width=0.7\textwidth]{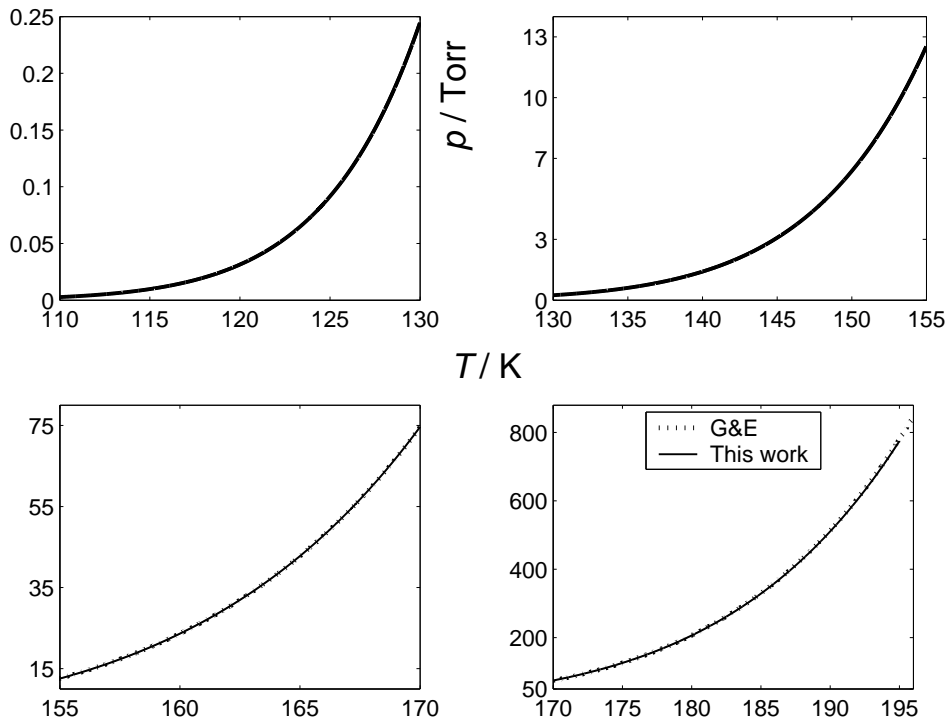}\\
  \caption{\footnotesize{The vapor pressure vs. the
temperature. Solid line: This work (TW) plotted for 110$\leq$$T
$$\leq$195, dotted line: G\&E plotted for 155$\leq$$T
$$\leq$196.}}\label{Fig1}
\end{figure}
\begin{figure}[h]
\centering
  \includegraphics[width=0.7\textwidth]{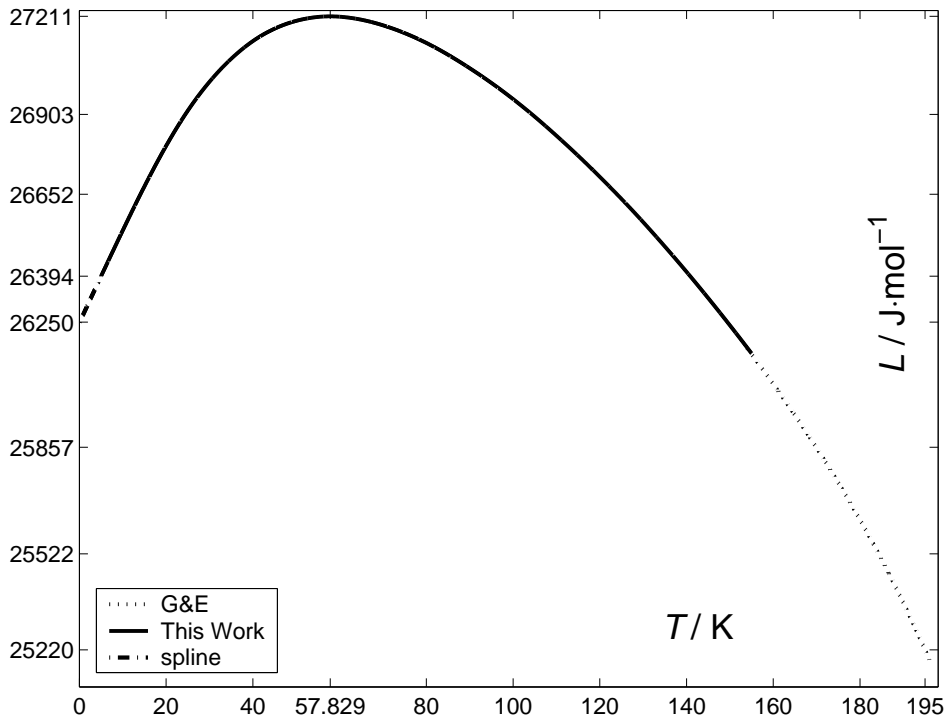}\\
  \caption{\footnotesize{The heat of sublimation vs. the
temperature. Solid line: This work (TW) plotted using the derived
Eq. (\ref{LTW}), dotted line: plotted using the same equation with
$p_{\,\rm G\&E}\,$, dash-dot line: arc of the spline through the
data ($T,L_{\,\rm TW}$) \& $T$=5$n$ K (0$\leq$$n$$\leq$31,
positive integer) shown in Table \ref{tab2}. This arc extrapolates the
solid line to temperatures below 5 K.}}\label{Fig2}
\end{figure}

\section{Results and Discussions}

\paragraph*{\textbf{Heat of sublimation at $\pmb{T=0}$.}}

Throughout this paper, we use the units and symbols recommended
by the \textit{International Union of Pure and Applied Chemistry}
(\textit{IUPAC}) \cite{IUPAC}. The energy is given in J and in cal
= 4.184 J, the pressure in Torr, and the temperature in K. Since
the original data were given in calories, we perform our
evaluations in this unit, taking $R$=1.98724 cal$
\cdot$K\textsuperscript{-1}$\cdot$mol\textsuperscript{-1}, then
convert the results to joules.

The G\&E heat capacity measurements, shown in the codes
(appendix), extend from 15.52 to 189.78 K. On such a large
interval there is no best equation that will represent the data
\cite{d3}. G\&E worked on a smooth curve through the data but did
not describe it. In order to represent the data, the alternative
is to subdivide the interval into sufficiently small intervals and
represent the data by a polynomial on each sub-interval in such a
way that the polynomial pieces blend smoothly making a spline
\cite{mat}.

MATLAB provides spline curve via the command
\verb"spline(x,y)" (see Appendix Section). It returns the piecewise
polynomial form of
the cubic spline interpolant with the not-a-knot end conditions,
having two continuous derivatives and breaks at all interior data
sites except for the leftmost and the rightmost one. The values of
the spline at the breaks \verb"spline(x,y,x(i))" coincide with the
data values \verb"y(i)". Cubic splines are more attractive for
interpolation purposes than higher-order polynomials \cite{mat}.

We will deal with molar physical quantities labeled by the
subscripts $s$ \& $g$ to differentiate between the solid and
gaseous phases. We denote by $L$ the latent heat of sublimation
and by $u_{i}$, $a_{i}$, $\mu_{i}$, $v_{i}$, $h_{i}$, $s_{i}$
($i$=$s,g$), the internal energy, free energy, chemical potential,
volume, enthalpy, entropy, respectively. We take the zero of
rotational energy to be that of the $J$=0 state and the zero of
vibrational energy to be that of the ground state, meaning that a
molecule at rest in the gas has an energy of zero at vanishing
temperature ($u_{g}(0)$=0). Let $\epsilon_{0}$ be the heat of
sublimation at $T$=0 which is, according to our energy convention,
the binding energy of the particles of the solid
($u_{s}(0)$=$a_{s}(0)
$=${h}_{s}(0)$=$\mu_{s}(0)$=$-\epsilon_{0}$$<$0).

The excellent agreement between the experimental \& spectroscopic
values of $s_{g}$ at 194.67 K is due to G\&E accurate measurements
and to the success of Debye's theory at low temperatures\footnote{
The more advanced theory elaborated in \cite{Suzuki}
reduces at low temperatures to Debye's theory.}. G\&E
used Debye's formula to evaluate $s_{s}$ for 0$\leq$$T$$\leq$15
K. However, they did
not explain their choice for Debye's temperature $\theta_{D}\,$.
In this work, the energy and entropy of the solid for temperatures
below 15.52 K are extrapolated by substitution of the Debye heat
capacity formula. Moreover, we will rely on Suzuki \& Schnepp's
assertion that the molar heat capacities of the solid carbon
dioxide ($c_{v}$ \& $c_{p}$) are equal within an error of
10$^{-5}$ per cent for such small temperatures \cite{Suzuki}.
Finally, we fix $\theta_{D}$ by equating the heat capacity due to
Debye with that measured by G\&E at 15.52 K (0.606
cal$\cdot$K\textsuperscript{-1}$\cdot$mol\textsuperscript{-1}).
Solving the equation using a CAS we find $\theta_{D}$=139.59 K.

The MATLAB codes provided in the appendix are split into three
parts. In Part (I), \verb"cd" represents the Debye heat
capacity. The vectors \verb"t" \& \verb"cp" show the temperature data
sites used by G\&E (15.52$\mapsto$189.78 K) and the corresponding measured
heat capacities (0.606$\mapsto$13.05
cal$\cdot$K\textsuperscript{-1}$\cdot$mol\textsuperscript{-1}),
respectively. These G\&E data sites are extended by the
temperature vector \verb"u" and the corresponding Debye heat
capacity vector \verb"v", respectively. The last two lines evaluate,
at the temperature vector \verb"Tn", the spline through the
extended data sites (\verb"t", \verb"cp"), the integrals
$\int_{0}^{T}c_{p}\,{\rm d}T'$=$h_{s}(T)+\epsilon_{0}$=$\Delta
h_{s}(T)$ (vector \verb"I") and $\int_{0}^{T}(c_{p}/T')\,{\rm
d}T'$=$s_{s}(T)$ (vector \verb"J"), with $T\in$ \verb"Tn".

The heat of sublimation $\epsilon_{0}$ is determined upon solving
the equation $\mu_{g}$=$\mu_{s}$ at any given temperature for
which the measured $L$ is known. The lead we had followed seeking
for higher accuracy led us to select the value of $L$=$6190$
cal$\cdot$mol\textsuperscript{-1} at $170$ K \cite[Eucken \&
Donath]{Meyers} \& \cite{Giauque}. We find $\epsilon_{0}$=$6273.4$
cal$\cdot$mol\textsuperscript{-1} and the calculation is shown
below.

With $\mu_{g}$=$a_{g}+p\,v_{g}$ \& $\mu_{s}$=${
h}_{s}-T\,s_{s}\,$, the equation $\mu_{g}$=$\mu_{s}$ reduces to
$\epsilon_{0}$=$\Delta h_{s}-T\,s_{s}-a_{g}-p\,v_{g}\,$. Upon
solving the Clapeyron equation for $p\,v_{g}$ we obtain
$p\,v_{g}$=$[L/(T\,{\rm d}\ln p/{\rm d}T)]+p\,v_{s}$, and finally
\begin{equation}\label{Eps}
    \epsilon_{0}=\Delta h_{s}-T\,s_{s}-a_{g}-\frac{L}
    {T({\rm d}\ln p/{\rm d}T)}-p\,v_{s}\,.
\end{equation}
We will make use of the G\&E empirical equation to evaluate
$p$ \& ${\rm d}\ln p/{\rm d}T$ at 170 K
\begin{equation}\label{GEp}
    p_{\,\rm G\&E}({\rm Torr})=10\,{\rm
    exp}[(a_{1}/T)+b_{1}+c_{1}\,T+d_{1}\,T^2]\quad (154\leq
    T\leq 196\,{\rm K}),
\end{equation}
($a_{1}$=$-$1354.210$\times$$\ln10$, $b_{1}$=8.69903$\times$$\ln
10$, $c_{1}$=0.001588$\times$$\ln 10$,
$d_{1}$=$-$4.5107$\times$$10^{-6}$$\times$$\ln 10$), and obtain
$p(170)$=74.59 Torr. Since $v_{s}$=25.55
cm\textsuperscript{3}$\cdot$mol\textsuperscript{-1} \cite{Suzuki},
the last term $p\,v_{s}$=0.06 cal$\cdot$mol\textsuperscript{-1} is
neglected. The term including $L$ equals
6190/(170$\times$0.108021)=337.08
cal$\cdot$mol\textsuperscript{-1}, and $\Delta h_{s}(170)\;\&
\;s_{s}(170)$ are the 85000\textsuperscript{th} components of the
vectors \verb"I" \& \verb"J": $\Delta h_{s}(170)-170\times
s_{s}(170)$=\verb"I(85000)-170*J(85000)"=$-$1227.8
cal$\cdot$mol\textsuperscript{-1}.

Now, we make our first hypothesis concerning the vapor. We assume
the validity of the first order virial expansion neglecting thus
the next terms, and this has always been the case for carbon
dioxide \cite{Giauque} at such low temperatures. We have then
\begin{equation}\label{v}
    p\,v_{g}=R\,T+B(T)\,p\,,
\end{equation}
thereby we can show that the term $a_{g}$ in (\ref{Eps}) is the
free energy of an ideal\footnote{In fact, we can show that the
correction for gas imperfection to $\mu_{g}$ is under the above
assumption $p\,v_{g}-R\,T$, implying $a_{g}$=$ a_{g\,\rm
ideal}\,$.} gas evaluated at the point $(T,p)$=$(170\,{\rm
K},74.59\,{\rm Torr})\,$. For the molecule of CO$_{2}$ we have
$a_{g}$=$a_{\rm t}+a_{\rm r}+a_{\rm v}$, which is the sum of the
translational, rotational and four vibrational contributions
$a_{\rm v}$=$2\,a_{\rm v1}+ a_{\rm v2}+a_{\rm v3}$ \cite{MQ,HG}.
With our choice of the origin of the energy, these contributions
write

\begin{equation}\label{fg}
\begin{tabular}{ll}
$a_{\rm t}=-R\,T\,\ln(\mathcal{C}\,{\rm e}\,T^{5/2}/p)\,;$ &
$\quad
a_{\rm r}=R\,\{-T\ln[T/(2\,\theta_{\rm r})]+\theta_{\rm r}/3\}\,;$ \\
$a_{{\rm v}i}=R\,T\ln[1-{\rm exp}(-\theta_{{\rm v}i}/T)]\,;$ &
\quad
$(T\geq 5 {\rm K})\;\&\;(i=1-3)$, \\
\end{tabular}
\end{equation}
with ${\mathcal C}$=7.575455$\times 10^5$ in SI units
(=$(2\,\pi\,m/h^2)^{3/2}\,k^{5/2}$) and $\theta_{\rm r}$=0.561,
$\theta_{\rm v1}$=954, $\theta_{\rm v2}$=1890, $\theta_{\rm
v3}$=3360 K. We have then $a_{g}(170)$=$-$7838.2
cal$\cdot$mol\textsuperscript{-1} leading with the previously
evaluated terms to $\epsilon_{0}$=6273.4
cal$\cdot$mol\textsuperscript{-1}.\\

\paragraph*{\textbf{Vapor pressure.}}
From now on we will assume $\epsilon_{0}$=6274
cal$\cdot$mol\textsuperscript{-1} (26250
J$\cdot$mol\textsuperscript{-1}). Upon substituting (\ref{v}) \&
(\ref{fg}) into $\epsilon_{0}$=$\Delta
h_{s}-T\,s_{s}-a_{g}-p\,v_{g}$ ($\mu_{g}$=$\mu_{s}$) and
rearranging the terms we obtain
\begin{equation}\label{pTW}
    p={\mathcal C}\,T^{5/2}Z_{\rm r}\,Z_{\rm v}\,
    {\rm exp}\{[\Delta h_{s}-T\,s_{s}-\epsilon_{0}-B(T)\,p]
    /RT\}\,,
\end{equation}
where $Z_{\rm v}$=$Z_{\rm v1}^{2}Z_{\rm v2}Z_{\rm v3}\,$, $Z_{\rm
vi}$=$1/[1-{\rm exp}(-\theta_{\rm vi}/T)]\,({\rm i}$=$1-3)$,
and\footnote{Because of the symmetry requirements of the total
wave function under the interchange of the two identical nuclei
\cite{MQ,HG}, $Z_{\rm r}$ is coupled with the nuclear partition
function and the above expression of $Z_{\rm r}$ no longer holds
for $T$ of the order of $\theta_{\rm r}\,$. However, as $T$
increases the separation of the two partition functions becomes
possible \cite{MQ}. The above formula for $Z_{\rm r}$ has been
derived using the Euler-MacLaurin expansion and can be used safely
for $T$ of the order of 5 K and higher values.} $Z_{\rm
r}$=$[T+\theta_{\rm r}/3]/(2\,\theta_{\rm r})\,$. Assuming that
$B(T)$ follows Berthelot's equation \cite{Giauque,d3}
\begin{equation}\label{B}
    B(T)\,p=R\,\ell_{1}[1-(\ell_{2}/T^{2})]\,p({\rm Torr})
\end{equation}
(where $\ell_{2}$=6$\times$304.1$^2$ K$^{2}$ and, in order to
express $B(T)\,p$ in cal$\cdot$mol\textsuperscript{-1}, we take
$\ell_{1}$=9$\times$304.1/(128$\times$72.8$\times$760) K/Torr), we
have solved numerically both equation (\ref{pTW}) and its
linearized form and the results coincide up to an insignificant
error. Upon substituting exp$[-B(T)p/RT]$=$1-B(T)p/RT$ into
(\ref{pTW}), the linearized equation yields
\begin{equation}\label{PTW}
p_{\,\rm TW}({\rm Torr})=\frac{p_{\,\rm
ideal}}{\{1+\ell_{1}[1-(\ell_{2}/T^{2})]p_{\,\rm ideal}/T\}} \quad
(T\geq 5\,{\rm K})\,,
\end{equation}
where $p_{\,\rm ideal}$ (in Torr) is the corresponding pressure
for an ideal gas
\begin{equation}\label{PI}
    p_{\,\rm ideal}({\rm Torr})=(760/101325)\,{\mathcal C}\,
    T^{5/2}Z_{\rm r}\,Z_{\rm v}\,{\rm exp}\{[\Delta h_{s}-
    T\,s_{s}-\epsilon_{0}]/RT\}\,.
\end{equation}

Table \ref{tab1} \& FIG. 1 compare values of the vapor pressure derived
in this work (TW) with those of G\&E (Eqs. (\ref{PTW}) \&
(\ref{GEp})). We have evaluated (\ref{GEp}) at temperatures below
the left-end point 154 K, as shown in Table \ref{tab1}, and the formula
remains applicable, however, for temperatures above 110 K; below
this temperature, equation (\ref{GEp}) diverges from (\ref{PTW}).
The third column (A) of Table \ref{tab1} shows values of the vapor pressure
evaluated using Antoine's equation \cite{d3}. The constants $
A_{1}$=6.81228, $B_{1}$=1301.679 \& $C_{1}$=$-$3.494 of Antoine's
equation have been evaluated by the National Institute of
Standards and Technology (NIST) \cite{nist} from G\&E data. The
equation writes
\begin{equation}\label{Ap}
    p_{\,\rm A}({\rm Torr})=(760/1.01325){\rm exp}\{\bar{A_{1}}-
    [\bar{B_{1}}/(A_{1}T+
    C_{1})]\}\quad (154.26\leq T\leq 195.89\,{\rm K}),
\end{equation}
where $\bar{A_{1}}$=$A_{1}\ln 10$, $\bar{B_{1}}$=$B_{1}\ln 10$.\\

{\footnotesize \topcaption{Vapor pressure data. The values of the
pressure shown in italics are evaluated at temperatures beyond the
assumed range of validity of the corresponding formula. The table
compares our results TW: This work (Eq. (\ref{PTW})) with those of
G\&E (Eq. (\ref{GEp})) \cite{Giauque} and Antoine's equation (Eq.
(\ref{Ap})) \cite{nist}. Nomenclature: NA=Not Applicable.
Conventions: 1) E--n=$10^{-n}\,$; 2) a letter C shown on the right
of a $p$-value indicates that a small correction for gas
imperfection has been added; if, otherwise, the values of $p$ with
and without correction are equal ($p$=$p_{\,\rm ideal}$). Since
the G\&E and A data are empirical, a letter C has been added to
all of them including those values evaluated beyond the assumed
range of validity.} \label{tab1}
\tablefirsthead{
  \hline
  \multicolumn{1}{l}{$T$/K} \hspace{0.17\textwidth} &
  \multicolumn{1}{l}{$p_{\,\rm TW}$/Torr} \hspace{0.17\textwidth} &
  \multicolumn{1}{l}{$p_{\,\rm G\&E}$/Torr} \hspace{0.17\textwidth} &
  \multicolumn{1}{l}{$p_{\,\rm A}$/Torr} \\ \hline}
\tablehead{
  \hline
  \multicolumn{4}{l}{\textsl{Continued from previous page}}\\
  \hline
  \multicolumn{1}{l}{$T$/K} \hspace{0.17\textwidth} &
  \multicolumn{1}{l}{$p_{\,\rm TW}$/Torr} \hspace{0.17\textwidth} &
  \multicolumn{1}{l}{$p_{\,\rm G\&E}$/Torr} \hspace{0.17\textwidth} &
  \multicolumn{1}{l}{$p_{\,\rm A}$/Torr} \\ \hline}
\tabletail{
  \hline
  \multicolumn{4}{r}{\textsl{Continued on next page}}\\
  \hline}
\tablelasttail{\hline}
\begin{supertabular*}{1.0\textwidth}{llll}
  65 & 3.4E--12 & NA & \textit{3.3E--12(C)} \\
  70 & 1.2E--10 & NA & \textit{1.3E--10(C)} \\
  75 & 2.8E--9 & NA & \textit{3.0E--9(C)} \\
  80 & 4.2E--8 & NA & \textit{4.7E--8(C)} \\
  85 & 4.7E--7 & NA & \textit{5.2E--7(C)} \\
  90 & 3.9E--6 & NA & \textit{4.3E--6(C)} \\
  95 & 2.6E--5 & NA & \textit{2.8E--5(C)} \\
 100 & 1.4E--4 & NA & \textit{1.5E--4(C)} \\
 105 & 6.8E--4 & NA & \textit{7.3E--4(C)} \\
 110 & 0.003 & \textit{0.003(C)} & \textit{0.003(C)} \\
 115 & 0.01 & \textit{0.01(C)} & \textit{0.01(C)} \\
 120 & 0.03 & \textit{0.03(C)} & \textit{0.03(C)} \\
 125 & 0.09 & \textit{0.09(C)} & \textit{0.09(C)} \\
 130 & 0.2 & \textit{0.2(C)} & \textit{0.2(C)} \\
 135 & 0.6 & \textit{0.6(C)} & \textit{0.6(C)} \\
 140 & 1.4 & \textit{1.4(C)} & \textit{1.4(C)} \\
 145 & 3.1 & \textit{3.1(C)} & \textit{3.1(C)} \\
 150 & 6.4 & \textit{6.4(C)} & \textit{6.3(C)} \\
 155 & 12.5 & 12.6(C) & 12.5(C) \\
 160 & 23.6 & 23.6(C) & 23.5(C) \\
 165 & 42.8(C) & 42.7(C) & 42.4(C) \\
 170 & 74.6(C) & 74.6(C) & 74.1(C) \\
 175 & 126(C) & 126(C) & 125(C) \\
 180 & 206(C) & 207(C) & 205(C) \\
 185 & 329(C) & 330(C) & 328(C) \\
 190 & 511(C) & 513(C) & 511(C) \\
 195 & 776(C) & 781(C) & 777(C) \\
\end{supertabular*}\\\\}

From Table \ref{tab1} we establish the following results. Equations
(\ref{GEp}) \& (\ref{Ap}) are still valid beyond their assumed
ranges of validity; the ranges are now extended right down below
their left-end points to include temperatures above 110 and 65 K,
respectively. Moreover, the vapor behaves as a polyatomic ideal
gas for temperatures below 155 K.

An instance of calculation is provided in the codes provided in
Part(II) of the appendix, which show the evaluation of the
ideal-gas pressure equation (\ref{PI}) \& the real-gas pressure
equation (\ref{PTW}) at 160, 180 \& 194.67 K. The evaluated
pressures are represented by the 3-vectors \verb"PI" and
\verb"PTW", respectively.\\

\paragraph*{\textbf{Heat of sublimation.}}
Combining different thermodynamic entities we establish the
equation
\begin{equation}\label{L}
    L(T)=\epsilon_{0}-\Delta h_{s}(T)+h_{g}(T)+
    [B(T)-(T\,{\rm d}B/{\rm d}T)]\,p(T)\,,
\end{equation}
where the last two terms add a correction for gas imperfection,
$p(T)$ is the vapor pressure and $h_{g}$ is the ideal-gas enthalpy
given by $h_{g}$=$R\,[(7T/2)$$-$$(\theta_{\rm r}/3)]
$$-$$T^{2}[{\rm d}(a_{\rm v}/T)/{\rm d}T]$ (Eq. (\ref{fg})).

Looking for extreme values we can first ignore the correction for
gas imperfection then justify it later. We have solved graphically
the equation ${\rm d}L/{\rm d}T$=0 ($c_{p\,s}$=$c_{p\,g}$) and
obtained the values 57.829 K for $T$ \& 6503.58
cal$\cdot$mol\textsuperscript{-1} for $L$ as shown in FIG. 2. We
will assume $L_{\rm max}$=6503.6 cal$\cdot$mol\textsuperscript{-1}
(27211 J$\cdot$mol\textsuperscript{-1}). Tables \ref{tab2} \& \ref{tab1}, however,
show that at 57.829 K the vapor behaves as an ideal gas, and this
justifies the omission of the correction terms in ${\rm d}L/{\rm
d}T$=0.

Substituting (\ref{B}) into (\ref{L}), this latter splits into two
equations whether we evaluate the vapor pressure using (\ref{GEp})
or (\ref{PTW})
\begin{eqnarray}
  L_{\,\rm G\&E} &=& \epsilon_{0}-\Delta h_{s}+
  h_{g}+R\,\ell_{1}[1-(3\ell_{2}/T^{2})]\,
  p_{\,\rm G\&E}\quad (154\leq T\leq 196\,{\rm K}),\label{GEL} \\
  L_{\,\rm TW} &=& \epsilon_{0}-\Delta h_{s}+
  h_{g}+R\,\ell_{1}[1-(3\ell_{2}/T^{2})]\,
  p_{\,\rm TW}\quad (T\geq 5 {\rm K})\,.\label{LTW}
\end{eqnarray}
Equations (\ref{GEL}) \& (\ref{LTW}) are plotted in FIG. 2. In
the codes provided in Part(III) of the appendix, we evaluate the
r.h.s of (\ref{LTW}) at 160, 180 \& 194.67 K (3-vector
\verb"LTW"). The value of the latent heat obtained at 194.67 K is
6030.4 cal$\cdot$mol\textsuperscript{-1} (25231
J$\cdot$mol\textsuperscript{-1}) or 6030.6
cal$\cdot$mol\textsuperscript{-1} (25232
J$\cdot$mol\textsuperscript{-1}) whether we calculate the r.h.s of
(\ref{LTW}) or (\ref{GEL}).\\

{\footnotesize \topcaption{Heat
of sublimation data. The values of the latent heat shown in
italics are evaluated at temperatures beyond the assumed range of
validity of the corresponding formula. The table compares our
results TW: This work (Eq. (\ref{LTW})) with those derived from
Eq. (\ref{GEL}) using G\&E pressure equation. Nomenclature: NA=Not
Applicable. Convention: a letter C shown  on the right of a
$L$-value indicates that a small correction for gas imperfection
has been added; if, otherwise, the values of $L$ with and without
correction are equal.} \label{tab2}
\tablefirsthead{
  \hline
  \multicolumn{1}{l}{$T$/K} \hspace{0.28\textwidth} &
  \multicolumn{1}{l}{$L_{\,\rm TW}$/J$\cdot$mol\textsuperscript{-1}} \hspace{0.28\textwidth} &
  \multicolumn{1}{l}{$L_{\,\rm G\&E}$/J$\cdot$mol\textsuperscript{-1}}
   \\ \hline}
\tablehead{
  \hline
  \multicolumn{3}{l}{\textsl{Continued from previous page}}\\
  \hline
  \multicolumn{1}{l}{$T$/K} \hspace{0.28\textwidth} &
  \multicolumn{1}{l}{$L_{\,\rm TW}$/J$\cdot$mol\textsuperscript{-1}} \hspace{0.28\textwidth} &
  \multicolumn{1}{l}{$L_{\,\rm G\&E}$/J$\cdot$mol\textsuperscript{-1}}
   \\ \hline}
\tabletail{
  \hline
  \multicolumn{3}{r}{\textsl{Continued on next page}}\\
  \hline}
\tablelasttail{\hline}
\begin{supertabular*}{1.0\textwidth}{lll}
  0 & 26250 & NA \\ 5 & 26394 & NA \\ 10 & 26538 & NA \\ 15 & 26676 & NA \\
  20 & 26804 & NA \\ 25 & 26914 & NA \\ 30 & 27005 & NA \\ 35 & 27077 & NA \\
  40 & 27133 & NA \\ 45 & 27172 & NA \\ 50 & 27197 & NA \\ 55 & 27209 & NA \\
  60 & 27210 & NA \\ 65 & 27201 & NA \\ 70 & 27183 & NA \\ 75 & 27158 & NA \\
  80 & 27128 & NA \\ 85 & 27091 & NA \\ 90 & 27048 & NA \\ 95 & 27002 & NA \\
  100 & 26951 & NA \\ 105 & 26896 & NA \\ 110 & 26836 & \textit{26836} \\
  115 & 26773 & \textit{26773} \\ 120 & 26707 & \textit{26707} \\
  125 & 26637 & \textit{26637} \\ 130 & 26565 & \textit{26565} \\
  135 & 26488 & \textit{26488} \\ 140 & 26408 & \textit{26408} \\
  145 & 26325 & \textit{26325} \\
  150 & 26239(C) & \textit{26239(C)} \\
  155 & 26149(C) & 26149(C) \\
  160 & 26055(C) & 26055(C) \\
  165 & 25958(C) & 25958(C) \\
  170 & 25855(C) & 25855(C) \\
  175 & 25745(C) & 25745(C) \\
  180 & 25629(C) & 25629(C) \\
  185 & 25504(C) & 25504(C) \\
  190 & 25368(C) & 25368(C) \\
  195 & 25221(C) & 25220(C) \\
  %\hline
\end{supertabular*}\\\\}

In concluding, it was of interest to further compare our results
for the pressure with those used by Stull \cite{d2} that, as
already stated, are less accurate than G\&E values. At
temperatures 138.8, 148.7, 153.6, 158.7 K, we read from \cite{d2}
the values 1, 5, 10, 20 Torr for the pressure, while our evaluated
values (Eq. (\ref{PTW})) are 1.16, 5.30, 10.42, 20.12 Torr,
respectively. Finally, values of the entropy of the solid at the
tabulated temperatures $T$=5$j$ K (1$\leq$$j$$\leq$39, positive
integer) form a sub-vector of \verb"J" and are obtainable upon
executing the codes \verb"q=2500:2500:97500; J(q)". For instance,
$s_{s}(160) $=\verb"J(80000)"=14.07,
$s_{s}(180)$=\verb"J(90000)"=15.50 and $s_{s}(194.67)
$=\verb"J(97335)"=16.52
cal$\cdot$K\textsuperscript{-1}$\cdot$mol\textsuperscript{-1}
(58.87, 64.85 \& 69.12
J$\cdot$K\textsuperscript{-1}$\cdot$mol\textsuperscript{-1},
respectively).

\section{Methods}

Concerning the numerical approach, given the accurate data for
the heat capacity at constant pressure of carbon dioxide and some
available data for the heat of sublimation, we employed the method
of splines to generate and evaluate a smooth curve representing
the heat capacity data. Dealing with a large number of data sites,
we preferred to use cubic splines, which are more attractive for
interpolation purposes than higher-order polynomials \cite{mat}.
Once the curve set, we proceeded to the evaluation of the change
of the enthalpy and entropy of the solid. The evaluation of the
relevant physical quantities concerning the vapor was rather
straightforward using almost fresh formulas from the thermodynamic
literature \cite{MQ,HG}. We used MATLAB to execute the task and
the calculated entities were used in subsequent vapor pressure and
heat of sublimation evaluations.

Now, concerning the theoretical approach, we mainly derived a
formula for the vapor pressure including a correction for gas
imperfection and effects for internal structure, as well as a
formula for the heat of sublimation with same purposes.\\

\section{acknowledgments}

\n The author acknowledges comments and suggestions by an
anonymous referee, which helped to improve the manuscript.\\

\n \textbf{Appendix}\\
\n This section is devoted to provide the main MATLAB codes, as a
part of the numerical method, leading to the results shown in this
paper.\\

\n \textit{\textbf{Part(I)}}\\
\n Part(I) shows the data sites used by G\&E (15.52$\mapsto$189.78 K)
\& (0.606$\mapsto$13.05 cal$\cdot$K\textsuperscript{-1}$\cdot
$mol\textsuperscript{-1}). We evaluate the spline through the
extended data sites (\verb"t", \verb"cp"),
the integrals $\int_{0}^{T}c_{p}\,{\rm d}T'$=$h_{s}(T)+\epsilon_{0}$=$\Delta
h_{s}(T)$ (vector \verb"I") and $\int_{0}^{T}(c_{p}/T')\,{\rm
d}T'$=$s_{s}(T)$ (vector \verb"J"), with $T\in$ \verb"Tn".
\begin{verbatim}
syms x z real;
f=(12/(x^3))*int((z^3)/(exp(z)-1),z,0,x);
g=(3*x)/(exp(x)-1); A=f-g; cd=3*1.98724*A;
u=0.01:0.01:15.25; xn=139.59./u;
v=real(double(subs(cd,x,xn))); t=[0 u 15.52 17.30
19.05 21.15 23.25 25.64 27.72 29.92 32.79 35.99
39.43 43.19 47.62 52.11 56.17 60.86 61.26 66.24
71.22 76.47 81.94 87.45 92.71 97.93 103.26 108.56
113.91 119.24 124.58 130.18 135.74 141.14 146.48
151.67 156.72 162.00 167.62 173.36 179.12 184.58
189.78]; cp=[0 v 0.606 0.825 1.081 1.419 1.791
2.266 2.676 3.069 3.555 4.063 4.603 5.195 5.794
6.326 6.765 7.269 7.302 7.707 8.047 8.370 8.703
8.984 9.189 9.421 9.671 9.893 10.07 10.27 10.44
10.69 10.88 11.08 11.27 11.45 11.64 11.84 12.07
12.32 12.57 12.82 13.05];
Tn=0.001:0.002:196.001; spcp=spline(t,cp,Tn);
I=0.002*cumsum(spcp); J=0.002*cumsum(spcp./Tn);
\end{verbatim}

\n \textit{\textbf{Part(II)}}\\
\n We evaluate the ideal-gas and real-gas pressures (Eqs. (\ref{PI}) \&
(\ref{PTW}))
at 160, 180 \& 194.67 K. The evaluated pressures are represented by the
3-vectors \verb"PI" and \verb"PTW", respectively.
\begin{verbatim}
Eps=6274; T=[159.999 179.999 194.669];
m=[80000 90000 97335]; ms=I(m)-(T.*J(m));
PC=7.575455*(10^5); l1=9*304.1/(128*72.8*760);
l2=6*(304.1^2); S=exp(ms./(1.98724*T));
Ztr=(1/(2*0.561))*((T.^(7/2)).*
(ones(size(T))+((0.561/3)./T)));
Zv=(1./((ones(size(T))-exp(-954./T)).^2)).*
(1./(ones(size(T))-exp(-1890./T))).*
(1./(ones(size(T))-exp(-3360./T)));
PI=((760/101325)*PC).*((Ztr.*Zv).*
(S.*exp(-Eps./(1.98742*T))));
V=(l1*((ones(size(T))-(l2./(T.^2))).*(PI./T)))+
ones(size(T)); PTW=PI./V;
T   =  160      180      194.67
PI  =  23.604   204.845  739.817
PTW =  23.632   206.308  754.942
\end{verbatim}

\n \textit{\textbf{Part(III)}}\\
\n We evaluate the the heat of sublimation (Eq. (\ref{LTW}))
160, 180 \& 194.67 K. The output is the 3-vector \verb"LTW".
\begin{verbatim}
IT=ones(size(T)); h1=954./(exp(954./T)-IT);
h2=1890./(exp(1890./T)-IT);
h3=3360./(exp(3360./T)-IT);
hv=1.98724*((2*h1)+h2+h3);
hg=((3.5*1.98724).*T)+hv-(((1.98724*0.561)/3)*IT);
GI=(1.98724*l1).*(IT-((3*l2)./(T.^2))).*PTW;
LTW=Eps-I(m)+hg+GI;
T            =  160        180        194.67
LTW(cal/mol) =  6227.4     6125.5     6030.4
LTW(J/mol)   =  26055      25629      25231
\end{verbatim}

\end{document}